\documentclass[aps,prl, twocolumn,showpacs,superscript address]{revtex4-2}
\usepackage{amsmath}
\usepackage{upgreek}
\usepackage{graphicx}
\usepackage[euler]{textgreek}
\usepackage[export]{adjustbox}
\usepackage{float}
\usepackage{physics}
\usepackage{mathrsfs}
\usepackage{bm}
\usepackage{caption}
\usepackage{textcomp}
\usepackage{natbib}
\usepackage{gensymb}
\usepackage{amsfonts} 
\usepackage{hyperref} 
\usepackage{xcolor}
\usepackage{blindtext}
\usepackage{multirow}

\hypersetup{
  colorlinks=true,
  linkcolor=cyan,
  urlcolor=blue,
  citecolor=cyan,
}

\graphicspath{ {.} }

\begin{document}
\captionsetup[figure]{labelsep=period,name={FIG.}, format={plain},justification={raggedright}}

\title{Ultrafast selective mid-infrared sublattice manipulation in the ferrimagnet FeCr\textsubscript{2}S\textsubscript{4}
}

\author{Davide Soranzio}
\email[e-mail:]{davideso@phys.ethz.ch}
\affiliation{Institute for Quantum Electronics, Eidgenössische Technische Hochschule (ETH) Zürich, CH-8093 Zurich, Switzerland}
\author{Matteo Savoini}
\affiliation{Institute for Quantum Electronics, Eidgenössische Technische Hochschule (ETH) Zürich, CH-8093 Zurich, Switzerland}
\author{Fabian Graf}
\affiliation{Institute for Quantum Electronics, Eidgenössische Technische Hochschule (ETH) Zürich, CH-8093 Zurich, Switzerland}
\author{Rafael T. Winkler}
\affiliation{Institute for Quantum Electronics, Eidgenössische Technische Hochschule (ETH) Zürich, CH-8093 Zurich, Switzerland}
\author{Abhishek Nag}
\affiliation{Center for Photon Science, Paul Scherrer Institute, 5232, Villigen-PSI, Switzerland}
\affiliation{Department of Physics, Indian Institute of Technology, Roorkee, Uttarakhand-247667, India} 
\author{Hiroki Ueda}
\affiliation{Center for Photon Science, Paul Scherrer Institute, 5232, Villigen-PSI, Switzerland}
\affiliation{Institute of Ion Beam Physics and Materials Research, Helmholtz-Zentrum Dresden-Rossendorf, Dresden 01328, Germany}
\author{Kenya Ohgushi}
\affiliation{Department of Physics, Graduate School of Science, Tohoku University, Sendai, Japan}
\author{Yoshinori Tokura}
\affiliation{RIKEN Center for Emergent Matter Science (CEMS), Wako, Japan}
\affiliation{Department of Applied Physics and Tokyo College, The University of Tokyo, Tokyo, Japan }
\author{Steven L. Johnson}
\email[e-mail:]{johnson@phys.ethz.ch}
\affiliation{Institute for Quantum Electronics, Eidgenössische Technische Hochschule (ETH) Zürich, CH-8093 Zurich, Switzerland}
\affiliation{Center for Photon Science, Paul Scherrer Institute, 5232, Villigen-PSI, Switzerland}

\begin{abstract}
FeCr\textsubscript{2}S\textsubscript{4} is a ferrimagnet with two oppositely ordered spin sublattices (Fe and Cr), connected via superexchange interaction, giving a non-zero net magnetic moment. 
We show, using time-resolved measurements of the magneto-optic Kerr effect, how the magnetic dynamics of the sublattices can be selectively manipulated by resonantly perturbing the Fe sublattice with ultrashort laser pulses. The mid-infrared excitation through intra-atomic Fe $d$-$d$ transitions triggers markedly slower dynamics in comparison to an off-resonant pumping affecting both of the two sublattices simultaneously. By changing probe wavelength to move in and out of resonance with the Fe $d$-$d$ transitions, we also show the specific contributions of the Fe sublattice to these dynamics.
\end{abstract}
\maketitle
\newpage


Ferrimagnets are an interesting class of magnetic materials that combine some features of antiferromagnets with those of ferromagnets. 
They are characterized by a finite net magnetic moment given by sublattices with opposite spin orientation but unequal magnetic moments.
The relative orientation of the magnetic sublattices is typically determined by the exchange interaction~\cite{Blundell2001}. This leads to intriguing potential applications in spintronics~\cite{Kim2022}, spin textures such as domain walls and skyrmions~\cite{Wu2020}, and in some cases the phenomenon of all-optical switching~\cite{Stanciu2007}. 
For light with photon energies near intra-atomic electronic transitions involving partially filled shells, strong element-specific enhancement of the magneto-optical Kerr effect (MOKE) response can be observed \cite{Ahrenkiel1973,Ohgushi2005,Ohgushi2008,Kocsis2018}.
In principle, exciting and monitoring the material resonantly to such transitions via light pulses in a time-resolved approach (TR-MOKE, \cite{Sato2022,Koopmans2003}) offers a unique pathway for selective manipulation of a single sublattice, provided that other contributions of mixed atomic character are distinctively separated in the energy spectrum.
Past studies have shown that resonant pumping can improve the efficiency of processes such as spin-wave excitation and switching, while resonant probing can help to distinguish the sublattice dynamics \cite{Khorsand2013,Stupakiewicz2017,Mikhaylovskiy2020,Mathias2012}. To our knowledge, however, no studies so far have reported how the individual sublattice dynamics vary as a result of resonant pumping.

\begin{figure}[t]
\includegraphics[width=\columnwidth]{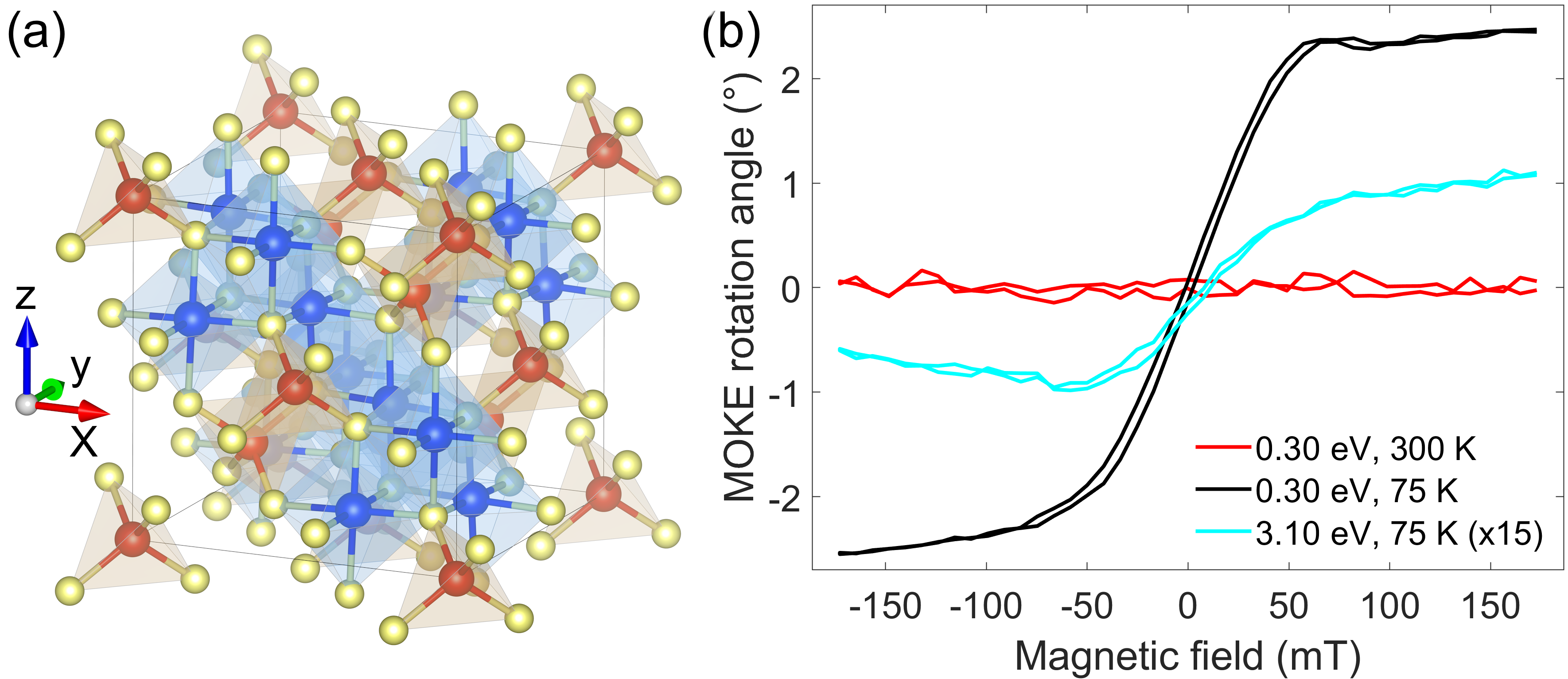}

\caption{Equilibrium properties of FeCr\textsubscript{2}S\textsubscript{4}. (a) Crystallographic cubic unit cell, where the Fe atoms are in red, Cr atoms are in blue and S atoms in yellow based on the data from Ref. \cite{Riedel1981} and plotted using VESTA \cite{Momma2011} (b) MOKE rotation hysteresis cycles for 0.30 eV (Fe $d$-$d$ transitions) and 3.10 eV (CT transitions) for $T$$<$$T_\mathrm{C}$$\approx$170 K; for comparison the hysteresis for $T$$>$$T_\mathrm{C}$ at 0.30 eV is also shown.} 

 \label{fig1}
\end{figure}

The ferrimagnet FeCr\textsubscript{2}S\textsubscript{4} is a semiconductor with a cubic spinel structure (space group Fd$\bar{3}$m) (Fig. \ref{fig1}(a)) \cite{Riedel1981,Chen1999,Tsurkan2005,Bertinshaw2014}, with some reports suggesting structural distortions below 65 K \cite{Deisenhofer2019,Maurer2003,Evans2022}. The iron cations (Fe\textsuperscript{2+}) are found in tetrahedral ($T_d$) sites, while the chromium ions (Cr\textsuperscript{3+}) are octahedrally ($O_h$) coordinated. A mixed valence configuration for the Fe and Cr cations was excluded by Chen et al. \cite{Chen1999}. Recently, the possibility of growing 2D-like crystals with analogous magnetic properties has been demonstrated, which makes the material attractive for applications in ultrathin devices \cite{Liu2024}. FeCr\textsubscript{2}S\textsubscript{4} becomes ferrimagnetic below the ordering temperature $T_\mathrm{C}$$\approx$170 K. This alignment is favored by the superexchange interaction between the half-filled Cr $t_{2g}$ orbitals and the partially-filled Fe $d$ orbitals, mediated by the sulfide S\textsuperscript{2-} anions \cite{Blundell2001,Sarkar2009,Chiuzbăian2017}. 
Polar MOKE studies on bulk samples have revealed a large, temperature-dependent MOKE rotation in the mid-infrared around photon energies of approximately 0.30 eV, enhanced by intra-atomic Fe $d$-$d$ transitions which become dipole-active due to the absence of local inversion symmetry at the iron site, leading to Davydov splitting \cite{Ohgushi2005,Ohgushi2008}. Differently, Cr $d$-$d$ transitions have a much smaller impact, comparable in magnitude to other non-resonant contributions, around 1.60 eV, at the edge of the visible range of the spectrum \cite{Kocsis2018,Ohgushi2008}. Around 3.10 eV, the MOKE rotation is about one order of magnitude smaller than in the mid-infrared, and mediated by charge-transfer (CT) transitions between S-3$p$ orbitals and mixed Fe-3$d$/Cr-3$d$ states \cite{Ogasawara2006,Sarkar2009,Ohgushi2005,Ohgushi2008}.
The strong enhancement of the magneto-optical coupling at photon energies in resonance with Fe $d$-$d$ transitions offers an attractive mechanism for both the manipulation and detection of the Fe sublattice. 
In this work, we report on polar TR-MOKE experiments on single-crystal FeCr\textsubscript{2}S\textsubscript{4} samples with pump and probe energies both on and off-resonance with respect to the intra-atomic Fe $d$-$d$ transitions. We find significant changes in the dynamics as a function of the wavelength of the pump pulse. In particular, we observe a dramatically slower response in the magnetic dynamics when the pump photon energy is tuned to resonance with the Fe $d$-$d$ transitions. Different probe energies allow us to analyze the roles of the sublattices in the dynamics. 

\begin{figure*}[t]
\includegraphics[width=\textwidth]{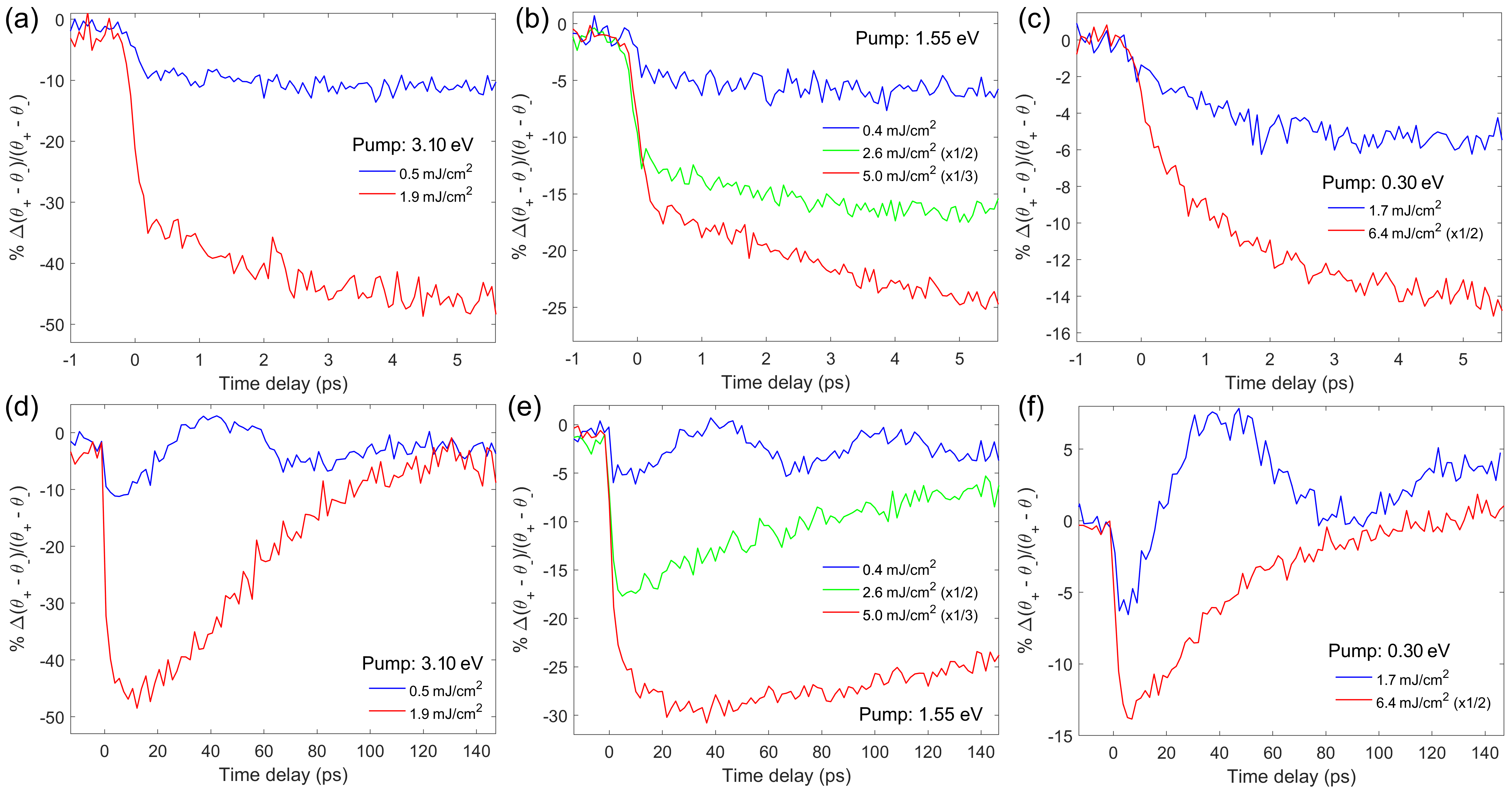}

\caption{TR-MOKE signal for FeCr\textsubscript{2}S\textsubscript{4} using a 3.10 eV probe. The 6 ps dynamics using (a) 3.10 eV (b) 1.55 eV and (c) 0.30 eV pump photon energies is reported in the top row. The bottom row presents the dynamics on a 150 ps time scale for (d) 3.10 eV (e) 1.55 eV and (f) 0.30 eV pump photon energies. All the measurements were acquired at $T$=75 K.}

 \label{fig2}
\end{figure*}

The single-crystal samples were grown using chemical vapor transport as reported by Ohgushi et al. \cite{Ohgushi2005}. All of our optical measurements were performed on the as-grown (111) surface. Although the easy axis corresponds to the $\langle100\rangle$ direction, mechanical polishing has been reported to cause a marked increase in the coercive field \cite{Ohgushi2005, Ohgushi2008}. This preferential alignment is connected to the temperature-dependent magnetocrystalline anisotropy energy, mainly related to the Fe sites \cite{Stapele1971}.
The experiments were performed using an amplified Ti:sapphire laser system (Coherent Astrella, photon energy 1.55 eV, pulse duration 100 fs, repetition rate 1 kHz) in combination with an optical parametric amplifier (TOPAS, Light Conversion). Mid-infrared pulses at 0.30 eV were achieved by difference-frequency generation of the signal and idler outputs of the optical parametric amplifier using a GaSe crystal, while visible 3.10 eV pulses were obtained by second-harmonic generation through a beta barium borate (BBO) crystal. More details on the experimental setup, list of the pump and probe spot sizes at the sample position, power reflectivity coefficients and intensity attenuation lengths are reported in the Supplemental Material (SM), Sects. I and II \cite{supplementary}.

\begin{figure*}[t]
\includegraphics[width=\textwidth]{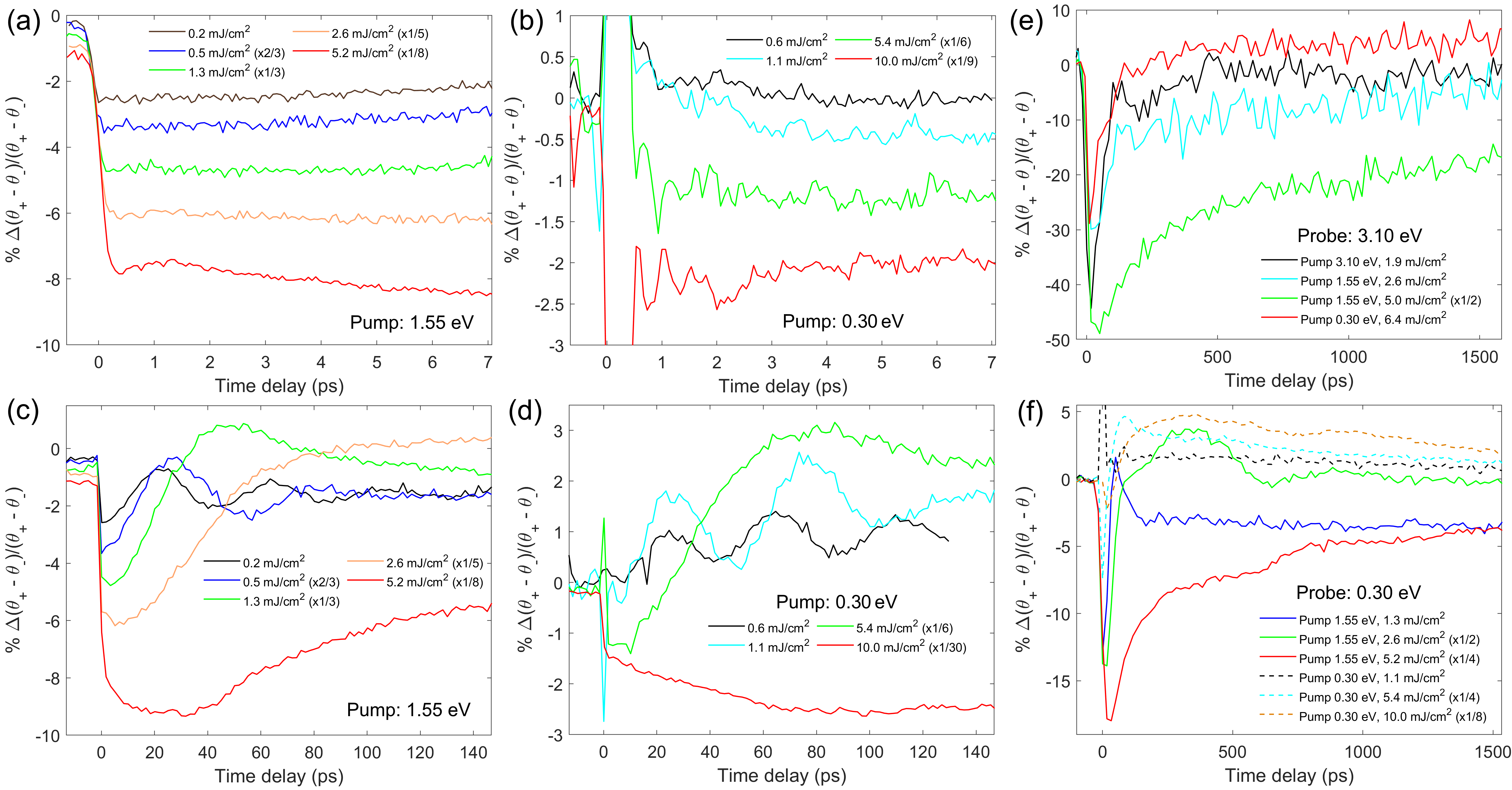}

\caption{TR-MOKE signal for FeCr\textsubscript{2}S\textsubscript{4} using a 0.30 eV probe. The 7 ps dynamics using (a) 1.55 eV and (b) 0.30 eV pump photon energies is reported in the top row. The bottom row presents the dynamics on a 150 ps time scale for (c) 1.55 eV and (d) 0.30 eV pump photon energies. The 1600 ps dynamics was studied using various pump photon energies and (e) 3.10 eV or (f) 0.30 eV probe photon energy. All the measurements were acquired at $T$=$75$~K.}

 \label{fig3}
\end{figure*}

In Fig. \ref{fig1}(b) we report the polar MOKE response from FeCr\textsubscript{2}S\textsubscript{4} at 0.30 eV and 3.10 eV through hysteresis cycles. The large few-degree mid-infrared MOKE rotation at 0.30 eV dwarfs the visible response at 3.10 eV at 75 K. The hysteresis vanishes above the critical temperature $T_\mathrm{C}$$\approx$170 K. These results are consistent with those reported by Ohgushi et al. \cite{Ohgushi2005}, where the enhanced response at 0.30~eV was attributed to the strong contributions of intra-atomic Fe $d$-$d$ transitions, whereas the smaller response at 3.10~eV was ascribed to contributions from interatomic CT transitions.
In the following, we focus on the time-resolved MOKE rotation dynamics, in particular on the out-of-equilibrium signal which is odd with respect to the externally applied magnetic field. To obtain it, we acquired pairs of datasets with a magnetic field of $\approx$170 mT, but with opposite directions with respect to the surface normal, namely $\theta_+$ and $\theta_-$. We express the signal as $\Delta(\theta_+ - \theta_-)/(\theta_+ - \theta_-)$.

We start by considering the case of a non-resonant probe (3.10 eV), where the magneto-optic effect is mediated by CT transitions. We explored three distinct time ranges with different time steps for selected incident fluence values (Figs. \ref{fig2}, \ref{fig3}(e)). Three pump photon energies were employed: 3.10 eV, 1.55 eV and 0.30 eV. In all these cases, the probe attenuation length is smaller than or equal to that of the pump.
The top row of Fig.~\ref{fig2} shows the initial dynamics over a 6 ps window around the temporal overlap between pump and probe pulses ($t$=0). In the non-resonant cases (3.10 eV and 1.55 eV, panels (a), (b)), a sudden drop occurs at time zero, followed by slower dynamics, which becomes more pronounced as the excitation energy increases. 
A major difference occurs once the pump photon energy is resonant with the Fe $d$-$d$ transitions (0.30 eV, panel (c)). Here, the dynamics do not exhibit a sharp drop at time zero, but rather a more gradual decrease. After this, the response then follows a trend similar to that observed for non-resonant pumping after about 2 ps. To describe the response after time zero, we use the empirical fit function $f(t)=G(t)\ast  \{\Theta(t)[A(1-e^{-t/\tau})+Bt+C]\}$, where $\Theta(t)$ is the Heaviside function, $G(t)$ is a Gaussian with unit area whose full width at half maximum (FWHM) is determined by the pump and probe pulse durations, $A$, $B$ and $C$ are fit parameters representing the relative amplitudes of the components of the response, and $\tau$ is a time constant. The time constants for the lowest and highest fluence values from each set of data among panels (a)-(c) are reported in Table \ref{table:1}.
For all the wavelength combinations, the FWHM of $G(t)$ was estimated to be shorter than 200 fs. More details on the fit results and pulse lengths can be found in the SM, Sect. I \cite{supplementary}.
In the bottom row of Fig. \ref{fig2}, we show the TR-MOKE rotation dynamics over a 150~ps range. Here, together with a gradual recovery MOKE signal after the excitation, we observe the emergence of a 24 GHz oscillation which disappears at high fluences for all the three pump photon energies.
In the longest time-delay window (1600 ps, Fig. \ref{fig3}(e)), the response is similar for all pump photon energies, consisting of a slow monotonic recovery after the initial excitation. As the pump fluence is increased, such as in the 1.55 eV case, the recovery in the first hundreds of picoseconds becomes at first slower, but then follows dynamics similar to that observed for the lower fluence.
\begin{table}[h]



\begin{tabular}{ |c|c|c| } 
\hline

\hline
 {Photon energy (eV)} & {Fluence (mJ/cm\textsuperscript{2})} & { Decay time $\tau$ (fs)} \\
 \hline
3.10 & 0.5 & $260\pm60$ \\ 
\hline
3.10 & 1.9 & $210\pm30$ \\ 
\hline
1.55 & 0.4 & $200\pm80$ \\ 
\hline
1.55 & 5.0 & $160\pm20$ \\ 
\hline
0.30 & 1.7 & $1600\pm500$ \\ 
\hline
0.30 & 6.4 & $890\pm60$ \\ 
\hline

\end{tabular}

\caption{Time decay constants derived using the fit function reported in the text for the lowest and highest investigated fluences from Fig. \ref{fig2}(a)-(c).}
\label{table:1}
\end{table}

We now turn our attention to the case of probe photon energy resonant to the Fe $d$-$d$ transitions (0.30 eV, Fig. \ref{fig3}). Again, three time ranges were collected, albeit only for two pump photon energies (1.55 eV and 0.30 eV). We chose not to acquire data for a pump photon energy of 3.10 eV since its attenuation length is approximately a factor of 10 smaller than at 0.30 eV. 
Starting from the 7 ps window shown in Fig. \ref{fig3} (a),(b), the response is predominantly flat after excitation.
For the 1.55 eV case, a small pre-pulse, arriving 19 ps before and estimated to be eight times smaller than the main pulse, was present in the pump. 
For degenerate pump and probe photon energies at 0.30 eV, it was not possible to fully remove the contribution of interference effects, which created coherent artifacts, localized mainly around the first few hundreds of fs after time zero (Fig. \ref{fig3}(b),(d),(f)) \cite{Vardeny1981,Cundiff2008}. Their identification is described in SM, Sect. III \cite{supplementary}.
Using the mid-infrared pump, although limited by the coherent artifacts at time zero, we do not observe a slow decrease of the out-of-equilibrium signal as in the case of the visible probe (Fig. \ref{fig2}(c)): after one picosecond the response is already flat and even presents a slow recovery at the highest fluences. 
Over the 150 ps range (bottom row, Fig. \ref{fig3}), we observe a response similar to that of the visible probe: a 24 GHz oscillation is present at the lower fluences, which gradually disappears, together with a strong signal decrease at the highest excitations. 
A comparison of the data shown in Fig. \ref{fig3} (c),(d), and (f) shows that the 24 GHz oscillation exhibits a gradual, strong, red-shift to around 2 GHz at the highest fluences.

One major contribution to the differences among the time dynamics at nearby fluences likely stems from the strong differences in attenuation length. 
This may explain, for example, why, for a 1.55 eV excitation with a given fluence, the 0.30 eV probe reveals dynamics which resembles those measured at lower fluences for the 3.10 eV probe (Figs. \ref{fig2}(e), \ref{fig3}(c)). A similar argument applies for the differences among distinct pump photon energies, but same probe, even after taking into account their reflectivities. 
However, the dynamics of a material is determined not simply by the absorbed energy, but also by the orbitals and quasi-particles involved.
An example is the TR-MOKE dynamics in Fig. \ref{fig2}(a)-(c). Considering the average absorbed excitation energy density, the highest fluence for the 3.10 eV pump in panel (a) corresponds to 9.8 \textmu J/m\textsuperscript{3}, whereas it is 4.3 \textmu J/m\textsuperscript{3} for 1.55 eV excitation in panel (b) and 2.2 \textmu J/m\textsuperscript{3} for 0.30 eV excitation in panel (c). Despite this, even decreasing the pulse energy by a factor of 10 for the 1.55 eV pump does not reproduce the slower dynamics seen for the 0.30 eV pump.
As shown for antiferromagnetic iron oxides \cite{Mikhaylovskiy2020}, resonant pumping of Fe $d$-$d$ transitions can weaken the magnetic interactions by altering the electronic configuration of the sublattices. This can transiently make a different alignment more favorable. The data of Fig. \ref{fig2}(a)-(c) show that, using a 3.10 eV probe to monitor the response sensitive to modifications that occur in both sublattices \cite{Sarkar2009,Ohgushi2005, Ohgushi2008,Ogasawara2006}, a 0.30 eV excitation leads to a slower TR-MOKE dynamics in comparison to other excitation photon energies that are not resonant with the intra-atomic Fe $d$-$d$ transitions. This dependence of the speed of the initial response on the pump photon energy does not occur when probing resonantly to the Fe sublattice (Fig. \ref{fig3}(a),(b)), where a fast drop was observed regardless of the selected pump.
This suggests that the slower response in Fig. \ref{fig2}(c) is connected to a change in the superexchange interaction between the excited Fe states and the Cr sublattice, triggered by the selective modification of the Fe electronic configuration. In fact, excitations at either 1.55 or 3.10 eV photon energies perturb directly both the two sublattices, not relying mainly on an indirect Cr spin reorganization via superexchange \cite{Sarkar2009, Ohgushi2008}. In the simplified semiclassical picture of the Heisenberg model, these pump-induced changes can be viewed as variations of the effective magnetic fields generated by each sublattice, which depend on the exchange interaction and spin orientation \cite{Blundell2001,Ogasawara2006}. 
Interaction with the phonon bath is expected to occur on a longer time scale \cite{Xu2021}.

The 24 GHz oscillation was previously reported by Ogasawara et al., and assigned to the precession of Cr spins \cite{Ogasawara2006}. This was motivated by the lack of oscillations at 0.5 eV probe photon energy (pump at 1.55 eV), ruling out a significant role of the Fe spins, limited by the relatively high magnetocrystalline anisotropy of this material \cite{Stapele1971}. Despite this, our data in Fig. \ref{fig3}(c),(d) clearly show oscillations at 24 GHz when probing resonantly to the Fe sublattice (0.30 eV). 
These experimental results can be reconciled by noting that a photon energy of 0.5 eV is at the very edge of the MOKE resonance associated with the Fe $d$-$d$ transitions \cite{Ohgushi2005,Ogasawara2006, Ohgushi2008}. Thus, MOKE data at this photon energy may have other contributions that are not specific to the Fe sublattice. Moreover, in Ogasawara et al. \cite{Ogasawara2006}, a slight oscillation appears to be in fact present in the datasets at 70 K and 90 K. The crystalline anisotropy, although predicted to be on the order of a few Tesla at 4 K \cite{Ogasawara2006}, is heavily influenced by temperature \cite{Stapele1971}. This is also evident from the coercive field growth while cooling \cite{Liu2024}. 
Therefore, we conclude that under our conditions ($T$=75 K), the magnetocrystalline anisotropy, while present, is considerably reduced. Since the oscillation frequency is close to that observed when probing at low fluences for 3.10 eV (Fig. \ref{fig2}(d)-(f)) and 1.9 eV \cite{Ogasawara2006}, we conclude that the two sublattices are coupled in their precession motion by the superexchange interaction.
No further oscillatory contributions were observed (SM, Sect. IV \cite{supplementary}).

Additional data on the FeCr\textsubscript{2}S\textsubscript{4} dynamics can be found in the SM \cite{supplementary}: out-of-equilibrium hysteresis curves (Sect. V), TR-MOKE ellipticity showing analogous dynamics and suggesting heating effects for the highest fluences, which give dramatic changes to the MOKE coefficients (Sect. VI), and time-resolved reflectivity data showing a non-negligible impact of the external magnetic field under $T_\mathrm{C}$ (Sect. VII).

In summary, we have shown that, by selectively pumping the Fe sublattice in the mid-infrared, we are able to modify dramatically the time scale of the TR-MOKE response connected to both the Fe and Cr sublattices, compared to non-resonant case. We propose that the possible mechanism consists of a distinct modification of the superexchange interaction, based on which electronic levels are perturbed by the pump.
Moreover, our work highlights the importance of resonant probing to distinguish the dynamics of the sublattices and to understand whether their spins are involved in specific collective modes. This approach can be extended to other magnetic materials provided that a clear spectral separation among the localized energy levels of the atoms is present.



\section*{Acknowledgments}
D.S. recognizes the support from the ETH Career Seed Fund 2023-1 through the project `Mid-infrared time-resolved magneto-optics'. This work was financially supported by JSPS KAKENHI Grant No. JP22H00102, and by JST CREST under Grant No. JP19198318. 

\section*{Data availability}
The dataset presented in this article is openly available \cite{dataset}.

\nocite{Soranzio2019,trebino2000,Klug2021,Checkhov2021}


\bibliography{biblio}

\end{document}


\setcounter{figure}{0}
\renewcommand{\figurename}{Fig.}
\renewcommand{\thefigure}{S\arabic{figure}}
\renewcommand{\thetable}{S\arabic{table}}
\title{Supplemental Material on Ultrafast selective mid-infrared sublattice manipulation in the ferrimagnet FeCr\textsubscript{2}S\textsubscript{4}
}

\author{Davide Soranzio}
\email[e-mail:]{davideso@phys.ethz.ch}
\affiliation{Institute for Quantum Electronics, Eidgenössische Technische Hochschule (ETH) Zürich, CH-8093 Zurich, Switzerland}
\author{Matteo Savoini}
\affiliation{Institute for Quantum Electronics, Eidgenössische Technische Hochschule (ETH) Zürich, CH-8093 Zurich, Switzerland}
\author{Fabian Graf}
\affiliation{Institute for Quantum Electronics, Eidgenössische Technische Hochschule (ETH) Zürich, CH-8093 Zurich, Switzerland}
\author{Rafael T. Winkler}
\affiliation{Institute for Quantum Electronics, Eidgenössische Technische Hochschule (ETH) Zürich, CH-8093 Zurich, Switzerland}
\author{Abhishek Nag}
\affiliation{Center for Photon Science, Paul Scherrer Institute, 5232, Villigen-PSI, Switzerland}
\affiliation{Department of Physics, Indian Institute of Technology, Roorkee, Uttarakhand-247667, India} 
\author{Hiroki Ueda}
\affiliation{Center for Photon Science, Paul Scherrer Institute, 5232, Villigen-PSI, Switzerland}
\affiliation{Institute of Ion Beam Physics and Materials Research, Helmholtz-Zentrum Dresden-Rossendorf, Dresden 01328, Germany}
\author{Kenya Ohgushi}
\affiliation{Department of Physics, Graduate School of Science, Tohoku University, Sendai, Japan}
\author{Yoshinori Tokura}
\affiliation{RIKEN Center for Emergent Matter Science (CEMS), Wako, Japan}
\affiliation{Department of Applied Physics and Tokyo College, The University of Tokyo, Tokyo, Japan }
\author{Steven L. Johnson}
\email[e-mail:]{johnson@phys.ethz.ch}
\affiliation{Institute for Quantum Electronics, Eidgenössische Technische Hochschule (ETH) Zürich, CH-8093 Zurich, Switzerland}
\affiliation{Center for Photon Science, Paul Scherrer Institute, 5232, Villigen-PSI, Switzerland}


\maketitle
\section{MOKE setup}
In all of the wavelength settings presented, we used an s-polarized pump beam arriving with a 2$^{\circ}$ incident angle to excite the system, and a p-polarized probe beam with a 7$^{\circ}$ angle with respect to the surface normal. A complete list of the pump and probe spot sizes at the sample position, power reflectivity coefficients and intensity attenuation lengths are reported in the Sect. A. 
To access the ferrimagnetic phase, we cooled the sample to 75~K using a cryostat with calcium fluoride windows, which had $\approx$93\% transmission over the entire investigated photon energy range. To orient the spins, we utilized a single-pole electromagnet, giving a DC magnetic field. The reflected probe beam was collected and sent to a half-wave plate and then split between two orthogonal polarizations using a Wollaston prism. The probe beams were finally recorded using Si or HgCdTe detectors depending on the photon energy. The MOKE angle was calculated as reported Sect. B. The cross-correlation between pump and probe pulse is discussed in Sect. C. Finally, in Sect. D we discuss the empirical data fitting model.

\subsection{Laser beam spot sizes at the sample}
In the experiment, we used multiple pump and probe photon energies. To implement the different configurations, the original 1.55 eV laser output was employed to generate the 0.30 eV and 3.10 eV photons through processes of difference-frequency and second-harmonic generation, respectively. These follow different paths in the setup, which can be interchanged through magnetic mounts. Moreover, different lenses were employed. This required an optimization to achieve similar spot sizes at the sample, among the pump beams as well as the probe beams. We report a summary of the spot sizes (full width at half maximum) for the various beams that were used during the measurements in Table \ref{tab1}.

\begin{table}[b]
\begin{center}
\small
\begin{tabular}{ |c|c|c| } 
\hline
Role &  Photon energy (eV) & Spot size (\textmu m\textsuperscript{2}) \\
 \hline
\hline
Pump & 0.30 & 440x420  \\ 
\hline
Pump & 1.55 & 417x367  \\ 
\hline
Pump & 3.10 & 237x584  \\ 
\hline
Probe & 0.30 & 263x249  \\ 
\hline
Probe & 3.10 & 168x192  \\ 
\hline

\end{tabular}
\caption{Spot sizes at the sample of the beams used during the experiment.}
\label{tab1}
\end{center}
\end{table}
\subsection{Determination of the rotation/ellipticity angles}
The rotation and ellipticity angles were obtained in the following way.
In absence of applied magnetic field, the reflected probe beam was directed towards a half- or quarter-wave plate. The resulting beam was then split between the horizontal and vertical polarizations, which were separately acquired by two distinct photodiodes.
The wave plate angle was set such that the intensity recorded by the two detectors was the same in the absence of applied magnetic field. Assuming an incoming beam whose polarization plane is parallel to the optical table of the laboratory, this corresponds to rotating the fast axis of the half-wave plate by a $\phi=\pi/8$ angle with respect to the horizontal. Similarly, the rotation angle was set to $\phi=\pi/4$ for the quarter-wave plate.

Using the Jones matrix formalism, it is possible to show that in the approximation of small angles, both the MOKE rotation and the ellipticity can be calculated using a simple formula.

The Jones matrix for the half-wave plate at $\phi=\pi/8$ is 
\begin{gather}
HW_{\pi/8}=
\begin{pmatrix}
\mathrm{cos}(\pi/8) & -\mathrm{sin}(\pi/8) \\
\mathrm{sin}(\pi/8) & \mathrm{cos}(\pi/8)
\end{pmatrix}
\begin{pmatrix}
1 & 0 \\
0 & -1 
\end{pmatrix}\begin{pmatrix}
\mathrm{cos}(\pi/8) & \mathrm{sin}(\pi/8) \\
-\mathrm{sin}(\pi/8) & \mathrm{cos}(\pi/8)
\end{pmatrix}=\frac{1}{\sqrt{2}}
\begin{pmatrix}
1 & 1 \\
1 & -1
\end{pmatrix}
\end{gather}
The Jones matrix for the quarter-wave plate at $\phi=\pi/4$ is 
\begin{gather}
QW_{\pi/4}=\frac{1}{2}e^{-i\pi/4}
\begin{pmatrix}
1 & -1 \\
1 & 1 
\end{pmatrix}
\begin{pmatrix}
1 & 0 \\
0 & i 
\end{pmatrix}\begin{pmatrix}
1 & 1 \\
-1 & 1 
\end{pmatrix}=\frac{1}{\sqrt{2}}
\begin{pmatrix}
1 & -i \\
-i & 1 
\end{pmatrix}
\end{gather}

We consider an electric field and express the ratio of its components as $\mathrm{tan}\left (\Phi \right )=E_y/E_x$, where $\Phi=\theta+i\eta$
\begin{gather}
\bm{E}=
\begin{pmatrix}
E_x\\
E_y  
\end{pmatrix}=E_x\begin{pmatrix}
1\\
\mathrm{tan}\left (\Phi \right ) 
\end{pmatrix}\approx E_x\begin{pmatrix}
1\\
\Phi
\end{pmatrix}
\end{gather}
where we assumed small angles $\Phi$. In fact, without any interaction with the sample, the beam was set to be polarized along the $x$-axis.

If the beam encounters the balanced half-wave plate, the resulting electric field is
\begin{gather}
M_H=\frac{1}{\sqrt{2}}
\begin{pmatrix}
1 & 1 \\
1 & -1
\end{pmatrix} E_x\begin{pmatrix}
1\\
\Phi
\end{pmatrix}=\frac{E_x}{\sqrt{2}}\begin{pmatrix}
1+\Phi\\
1-\Phi
\end{pmatrix}
\end{gather}
The beam intensities, which we consider to be proportional to the recorder voltage values from the photodiodes through the same constant P, along the two axes are 
\begin{equation}
I_{Hx}=P E_x^2|(1+\Phi)|^2/2=E_x^2|(1+\theta+i\eta)|^2/2=E_x^2[(1+\theta)^2+\eta^2]/2\\
\end{equation}
\begin{equation}
I_{Hy}=P E_x^2|(1-\Phi)|^2/2=E_x^2|(1-\theta-i\eta)|^2/2=E_x^2[(1-\theta)^2+\eta^2]/2\\
\end{equation}
leading to
\begin{equation}
\theta=\frac{1}{2}\frac{I_{Hx}-I_{Hy}}{I_{Hx}+I_{Hy}}    
\end{equation}

Similarly, in the case of the balanced quarter-wave plate, the resulting electric field is

\begin{gather}
M_Q=\frac{1}{\sqrt{2}}
\begin{pmatrix}
1 & -i \\
-i & 1
\end{pmatrix} E_x\begin{pmatrix}
1\\
\Phi
\end{pmatrix}=\frac{E_x}{\sqrt{2}}\begin{pmatrix}
1-i\Phi\\
-i+\Phi
\end{pmatrix}
\end{gather}
The beam intensities, proportional to the recorder voltage values from the photodiodes, along the two axes are 
\begin{equation}
I_{Hx}=P E_x^2|(1-i\Phi)|^2/2=PE_x^2|(1-i\theta+\eta)|^2/2=PE_x^2[(1+\eta)^2+\theta^2]/2\\
\end{equation}
\begin{equation}
I_{Hy}=P E_x^2|(-i+\Phi)|^2/2=PE_x^2|(-i+\theta+i\eta)|^2/2=PE_x^2[\theta^2+(1-\eta)^2]/2\\
\end{equation}
leading to
\begin{equation}
\eta=\frac{1}{2}\frac{I_{Qx}-I_{Qy}}{I_{Qx}+I_{Qy}}    
\end{equation}

\subsection{Cross-correlation between pump and probe pulses}
\begin{figure}[h]
\includegraphics[width=16 cm]{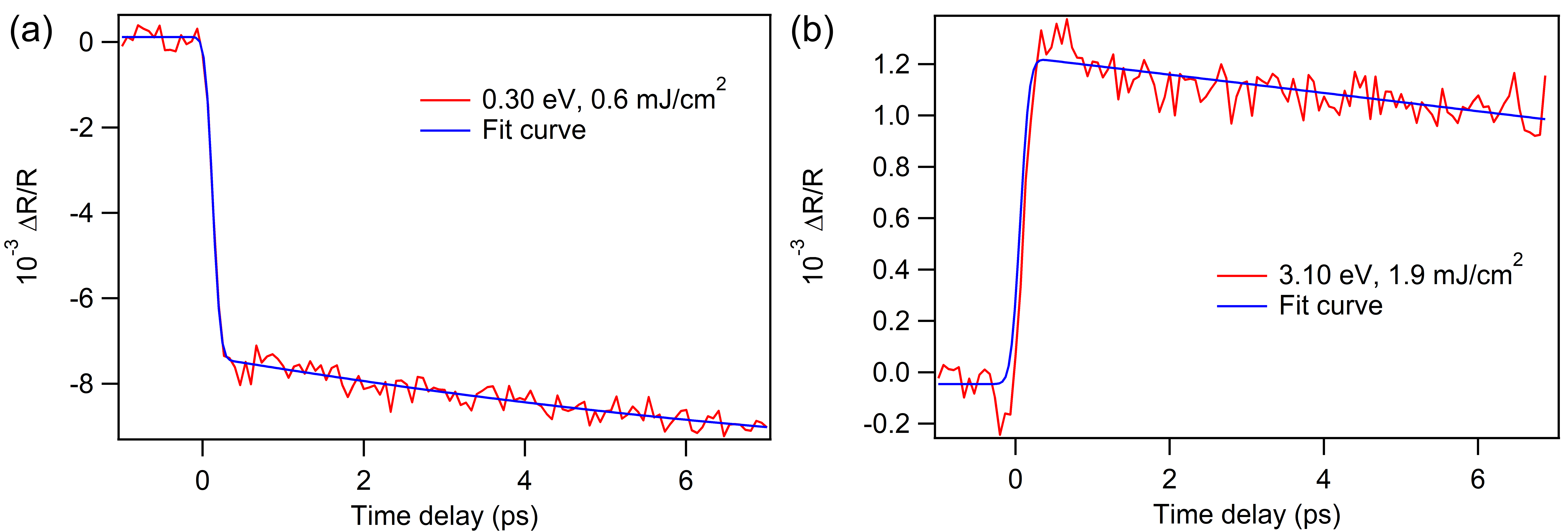}

\caption{Time-resolved reflectivity of the horizontal component in the (a) 0.30 eV pump - 0.30 eV probe (b) 3.10 eV pump - 3.10 eV probe wavelength combinations. The scans were recorded at $T$=75 K.}
\label{sfig1}
\end{figure}

We estimated the full width at half maximum (FWHM) temporal overlap between pump and probe pulse durations through the cross-correlation between pump and probe 0.30 eV mid-infrared pulses as follows.
We took a single-diode contribution coming from the TR-MOKE scans in Fig. 3(b) under a 0.6 mJ/cm\textsuperscript{2} fluence and plotted in Fig. \ref{sfig1}(a) the dynamical change in reflectivity for the horizontal component of the beam arriving at the Wollaston polarizer. As it can be seen from the graph, the main drop is almost immediate on such time scale. We fitted the data near time zero using the expression (1) from Soranzio et al. \cite{Soranzio2019}, approximating the response after time zero as the following convolution.

\begin{equation}
\Delta R/R(t)=G(t)\ast \left[\Theta(t)\left( \sum_{n=1}^{2}  A_ne^{-t/\tau_n}+D \right ) \right]
\end{equation}

where $\Theta(t)$ is the Heaviside function, $G(t)$ is a Gaussian with unit area, whose FWHM is determined by the pump and probe pulse durations the cross-correlation between the pulses (approximated as Gaussian), $A_n$ are amplitudes, $\tau_n$ are time-decay constants and $D$ is a constant representing long-lasting heating effects.
We obtained a FWHM of (180\textpm10) fs for the cross-correlation Gaussian. A more accurate estimation of the single pulse length could be obtained using autocorrelator methods. Repeating the same procedure for the pump-probe degenerate 3.10 eV configuration (Fig. \ref{sfig1}(b)) under a 1.9 mJ/cm\textsuperscript{2} fluence, we derive a FWHM for the cross-correlation Gaussian of (180\textpm40) fs.
Therefore, the difference in the dynamics near time zero among the three pump photon energies (Fig. 2(a)-(c)) cannot be explained by a much longer mid-infrared pulse at 0.30 eV compared to the non-resonant pump photon energies.

\subsection{Estimation of time constants}
To compare the decrease in the TR-MOKE signal shortly after time zero among the different wavelength datasets, we used an empirical fit function.  
\begin{equation}
f(t)=G(t)\ast\{\Theta(t)[A(1-e^{-t/\tau})+Bt+C]\}
\end{equation}
where $\Theta(t)$ is the Heaviside function, $G(t)$ is a Gaussian with unit area, whose full width at half maximum (FWHM) is determined by the pump and probe pulse durations, $A$ and $B$ are amplitudes, $\tau$ is the time decay constant and $C$ is a constant offset. In $f(t)$, having $G(t)$ described as a Gaussian function, is convolved with the response of the material.

The FWHM values have been estimated using a FROG characterization \cite{trebino2000} of the pulse duration for the fundamental 1.55 eV beam (around 110 fs) or by fitting experimental reflectivity data using expression (1) from Soranzio et al. \cite{Soranzio2019}, as discussed in the previous subsection.
In Table \ref{tab2} we report the results obtained from the data in Figs. 2 and 3 of the main text. The 0.30 eV pump - 0.30 eV probe configuration was not considered for the fitting analysis due to the presence of coherent artifacts at time zero; see Fig. 3(b) and related discussion in the main text \cite{Vardeny1981,Cundiff2008}.
A more advanced model would include a fully ab-initio description of the underlying phenomena as well as a direct experimental measurement of each pulse duration. 

\begin{table}[h]
\begin{center}
\small
\begin{tabular}{ |c|c|c|c| } 
\hline
Pump photon energy (eV) &  Probe photon energy (eV) & Fluence (mJ/cm\textsuperscript{2}) & Time constant (fs) \\
 \hline
\hline
3.10 & 3.10 & 0.5 & 260\textpm60  \\ 
\hline
3.10 & 3.10 & 1.9 & 210\textpm30 \\ 
\hline
1.55 & 3.10 & 0.4 & 200\textpm80 \\ 
\hline
1.55 & 3.10 & 2.6 & 190\textpm30  \\ 
\hline
1.55 & 3.10 & 5.0 & 160\textpm20 \\ 
\hline
3.10 & 3.10 & 1.7 & 1600\textpm500   \\ 
\hline
3.10 & 3.10 & 6.4 & 890\textpm60   \\ 
\hline
1.55 & 0.30 & 0.2 & 8\textpm5 \\ 
\hline
1.55 & 0.30 & 0.5 & 60\textpm20 \\ 
\hline
1.55 & 0.30 & 1.3 & 50\textpm10 \\ 
\hline
1.55 & 0.30 & 2.6 & 53\textpm8 \\ 
\hline
1.55 & 0.30 & 5.2 & 55\textpm7 \\ 
\hline

\end{tabular}
\caption{Time-decay constants obtained from the TR-MOKE data sets in Figs. 2 and 3 of the Main Text.}
\label{tab2}
\end{center}
\end{table}

\section{Equilibrium optical properties}
\begin{figure}[h]
\includegraphics[width=\textwidth]{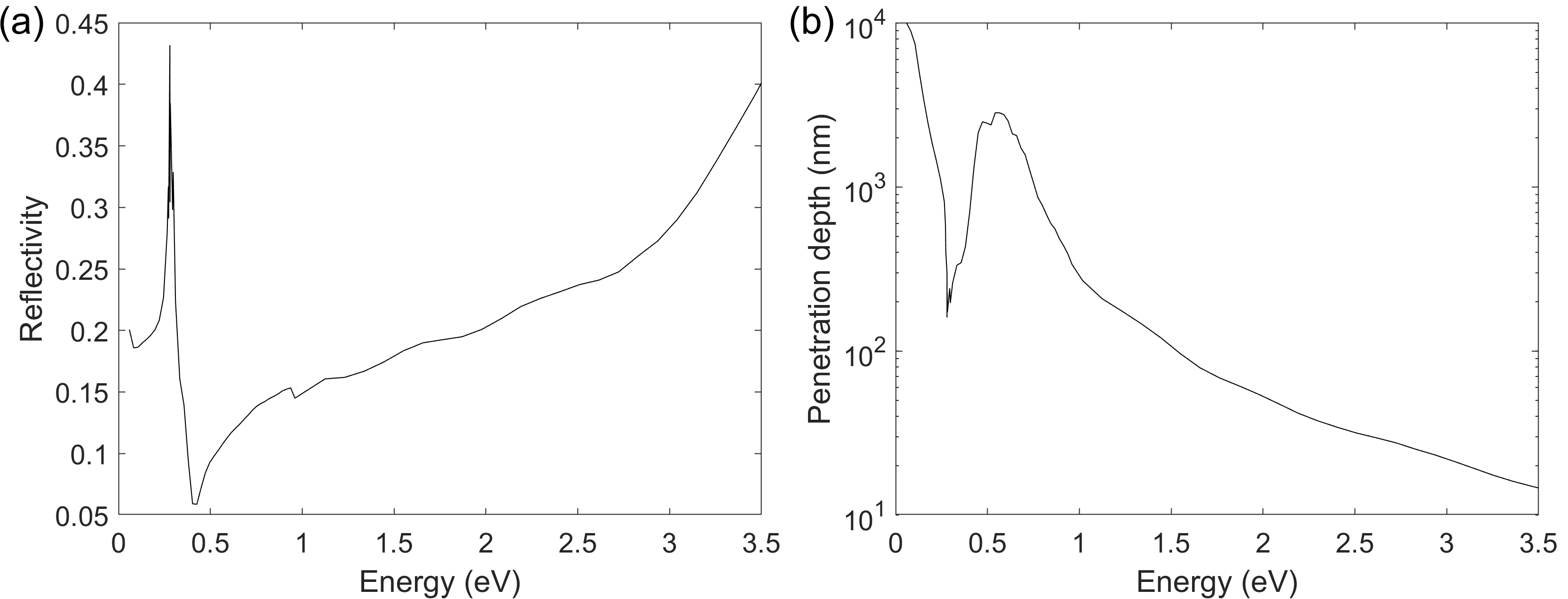}

\caption{(a) Power reflectivity and (b) penetration depth in FeCr\textsubscript{2}S\textsubscript{4} as a function of the photon energy. The curves were obtained using the dielectric function data reported by Ogasawara et al. \cite{Ogasawara2006}.}
\label{sfig2}
\end{figure}

Starting from the imaginary part of the diagonal element of the dielectric function reported by Ogasawara et al. \cite{Ogasawara2006}, we report the equilibrium reflectivity and intensity attenuation length as a function of the photon energy at $T$=6 K (Fig. \ref{sfig2}).
The power reflectivity coefficients at 0.30, 1.55, and 3.10 eV photon energies result to be 0.33, 0.18 and 0.30, respectively.  The corresponding intensity attenuation lengths are 197, 96, and 20 nm, respectively.

In spite of changes in the nature of the ferrimagnetic phase \cite{Bertinshaw2014} below $T$=60 K, we do not expect marked changes in the equilibrium response given the small changes in the lattice parameters and magnetization, as it was previously reported for the analogous spinel ferrimagnetic compound CoCr\textsubscript{2}S\textsubscript{4} \cite{Bertinshaw2014,Ohgushi2008}. Larger variations are expected above the critical temperature $T_C$=170 K, for example at room temperature as reported by Ohgushi et al. \cite{Ohgushi2008}.

The properties were derived using the Kramers-Kronig relations considering normal incidence and in the condition of isotropical response, to which we approximate our high-symmetry (space group Fd$\bar{3}m$) cubic crystal.

Although the reflectivity remains around values of 0.2-0.3 for most of the energy range, the penetration depth presents a strong energy dependence. In particular, comparing the photon energies we employed during the experiment (0.30, 1.55 and 3.10 eV), at the mid-infrared photon energy the penetration depth is about 2 times larger than the near-infrared one and 10 times than the visible one.
This is an important point, making it more appropriate to compare the pump impact in terms of excitation density (energy per volume) rather than fluences for the different pump photon energies.
\clearpage
\section{Coherent artifacts for the degenerate 0.30 eV configuration}

\begin{figure}[h]
\includegraphics[width=\textwidth]{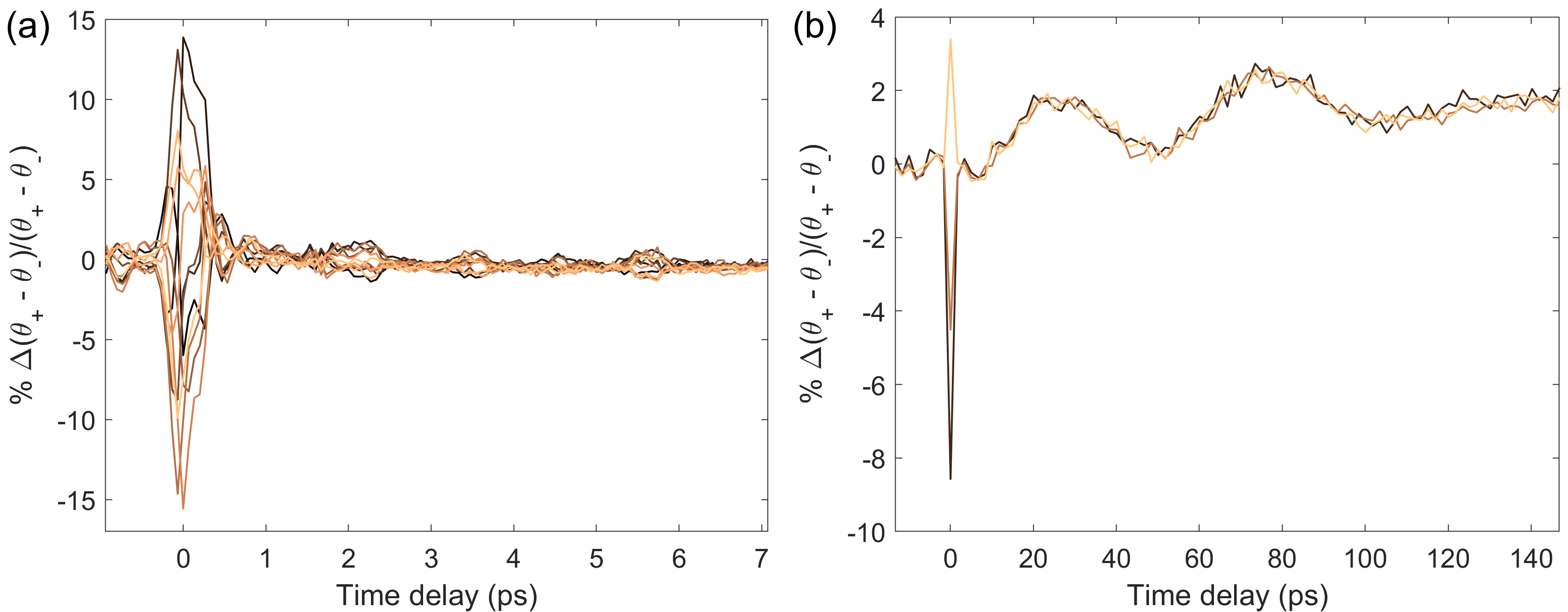}

\caption{Coherent artifacts in the dynamics for the pump 0.30 eV - probe 0.30 eV configuration. (a) 7 ps (b) 150 ps windows. The different colors represent subsequent scans. The scans were recorded at $T$=75 K under a 1.1 mJ/cm\textsuperscript{2} incident fluence.}
\label{sfig3}
\end{figure}

For the degenerate 0.30 eV configuration, it was not possible to fully remove the coherent artifacts from the pump in the acquired data \cite{Vardeny1981,Cundiff2008}. This alters the recorded response, which is not exactly reproduced in subsequent scans due to fluctuations in the pump and probe pulses structures. In Fig. \ref{sfig3}, we report two time ranges to visualize such effects. In panel (a), close to time zero, the coherent effects are more pronounced with a major peak in correspondence of the pulse overlap, followed by minor oscillations. Longer ranges using a coarser sampling like the one in panel (b) are only affected for a pair of points, again at time zero.

\section{Out-of-equilibrium oscillatory response}
\begin{figure}[h]
\includegraphics[width=\textwidth]{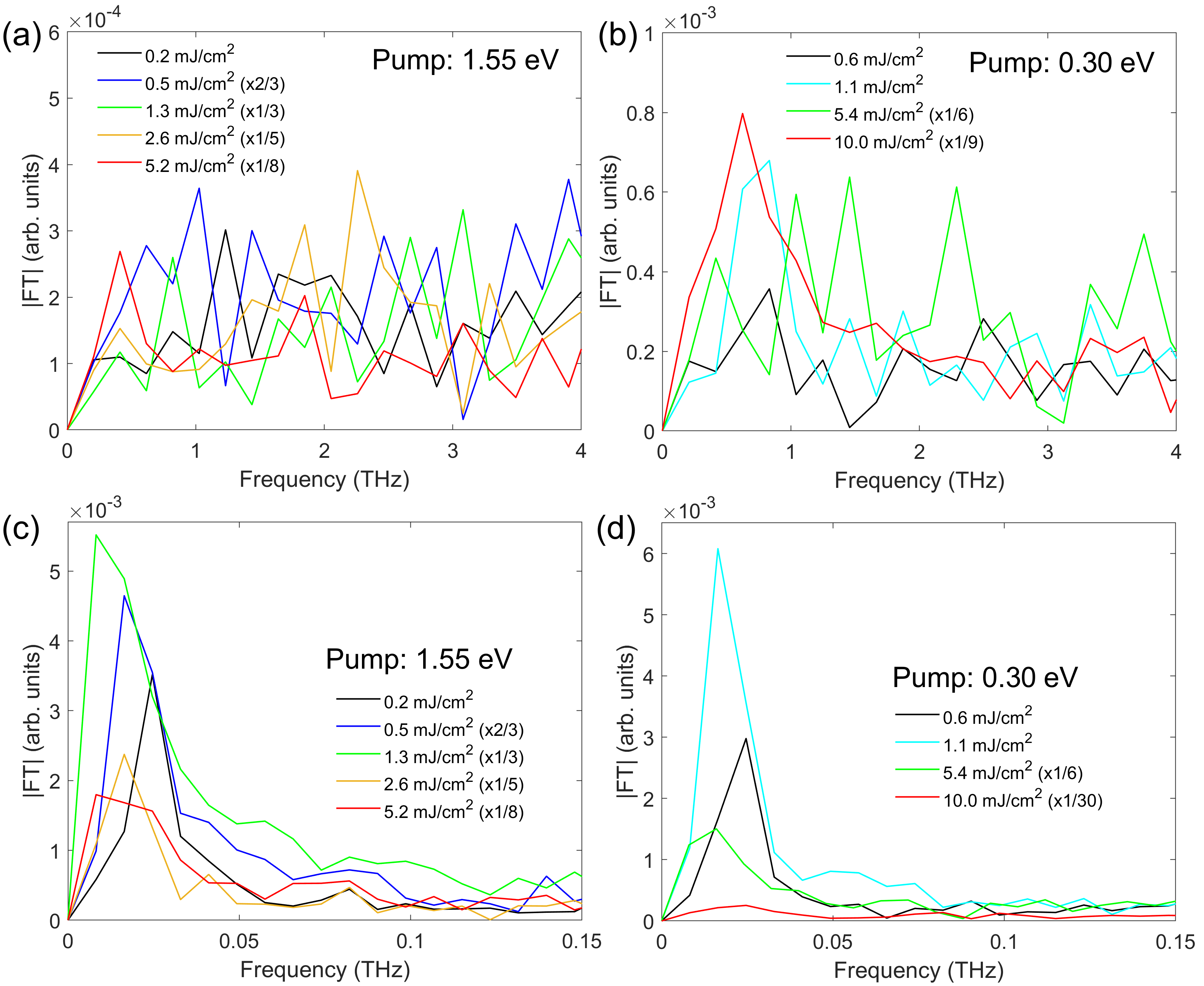}

\caption{Fourier transform of the oscillatory response in Fig. 3 (0.30 eV probe photon energy) of the Main Text after subtracting a quadratic background in the time ranges (a) 1.2-6.0 ps, (b) 1.2-6.0 ps, (c) 10-130 ps and (d) 10-130 ps.}
\label{sfig4}

\end{figure}

In Fig. \ref{sfig4}, we report in panels (a)-(d), the Fourier-transform results of the curves to the corresponding (a)-(d) panels in Fig. 3 of the Main Text. Before taking the Fourier transform, we subtracted a non-oscillatory second-order polynomial background.

\begin{figure}[h]
\includegraphics[width=\textwidth]{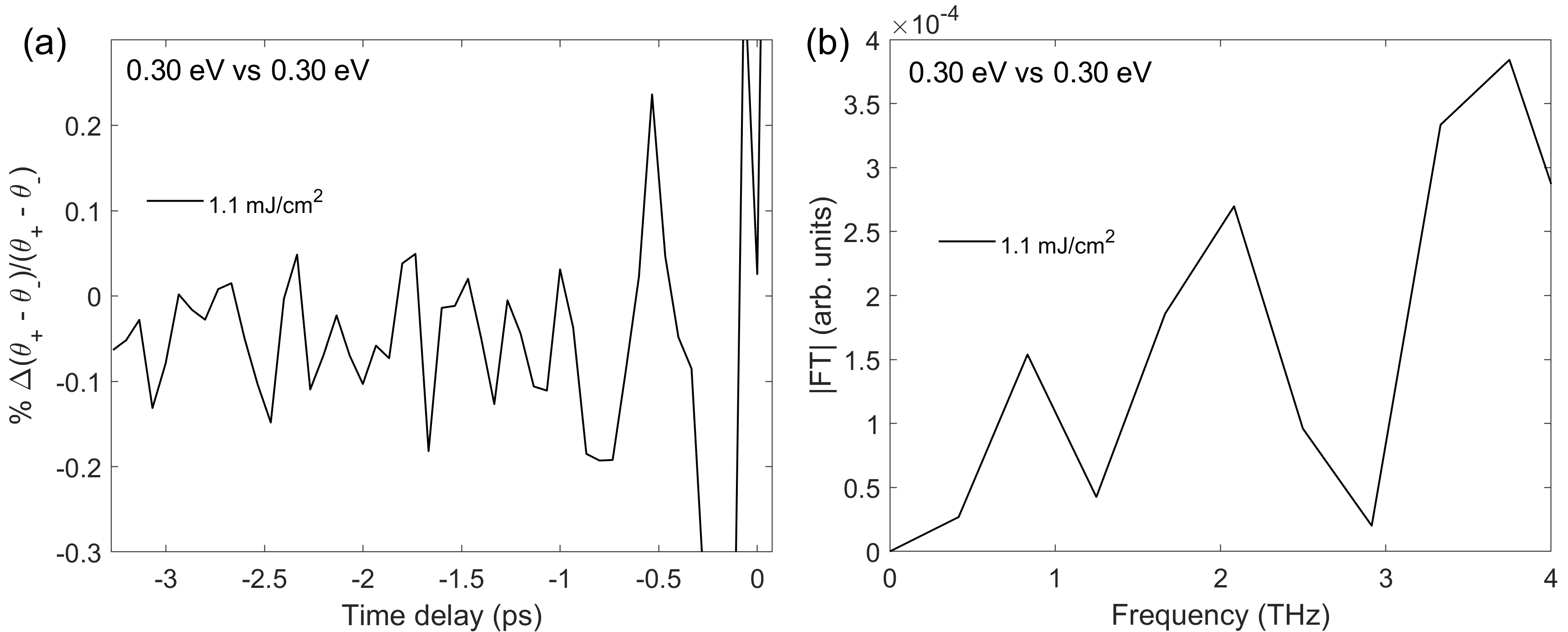}

\caption{Time-resolved MOKE signal for FeCr\textsubscript{2}S\textsubscript{4} using a 0.30 eV pump and 0.30 probe photon energy before time zero. (a) Time domain (b) Frequency domain obtained through a Fourier transform of the data between -3.3 and -0.9 ps after subtracting a quadratic background. The scan was acquired at $T$=75 K.}
\label{sfig5}

\end{figure}

The top row shows the Fourier transform of time ranges close to time zero. We observe in panel (b) a possible indication for a faster mode close to 1 THz for most of the collected fluences. With respect to the signal before time zero (Fig. \ref{sfig5}), we see a different spectrum.
However, for 5.4 mJ/cm\textsuperscript{2}, this peak does not emerge from the background, thus not supporting this hypothesis and suggesting it to be an artifact. As shown in the last section, this most likely results from some residual coherent artifacts (Fig. \ref{sfig3}(a)). In the case of a non-resonant driving (Fig. \ref{sfig4}(a)), we did not observe it.

In the bottom row, we report the results for the intermediate time window, where a slow oscillation corresponding to the spin precession appears.
For the three lowest fluences in panel (c), corresponding to time traces with at least one oscillation period in the acquired time window, we obtained a  peak at (24\textpm4) GHz (0.2 mJ/cm\textsuperscript{2}), (16\textpm4) GHz (0.5 mJ/cm\textsuperscript{2}) and (8\textpm4) GHz (1.3 mJ/cm\textsuperscript{2}) revealing a clear red-shift as the fluence is increased.

Regarding panel (d), only the two lowest fluences present one or more oscillation periods in the time domain window. In this case, the results are (24\textpm4) GHz (0.6 mJ/cm\textsuperscript{2}), (16\textpm4) GHz (1.1 mJ/cm\textsuperscript{2}).

\section{Pumped MOKE Rotation Hysteresis curves}

\begin{figure}[h]
\includegraphics[width=\textwidth]{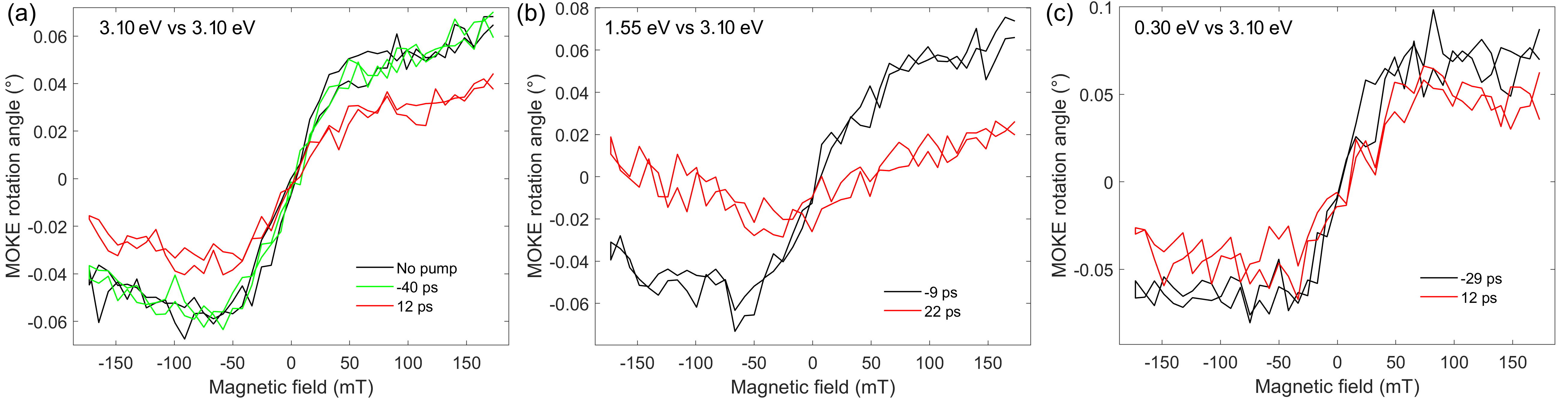}

\caption{Pumped MOKE rotation curved at different time delays for 3.10 eV probe photon energy, using (a) 3.10 eV (1.9 mJ/cm\textsuperscript{2}), (b) 1.55 eV (5.0 mJ/cm\textsuperscript{2}) and (c) 0.30 eV pump photon energies (6.4 mJ/cm\textsuperscript{2}). All the measurements were acquired at $T$=75 K.}
\label{sfig6}
\end{figure}

\begin{figure}[h]
\includegraphics[width=\textwidth]{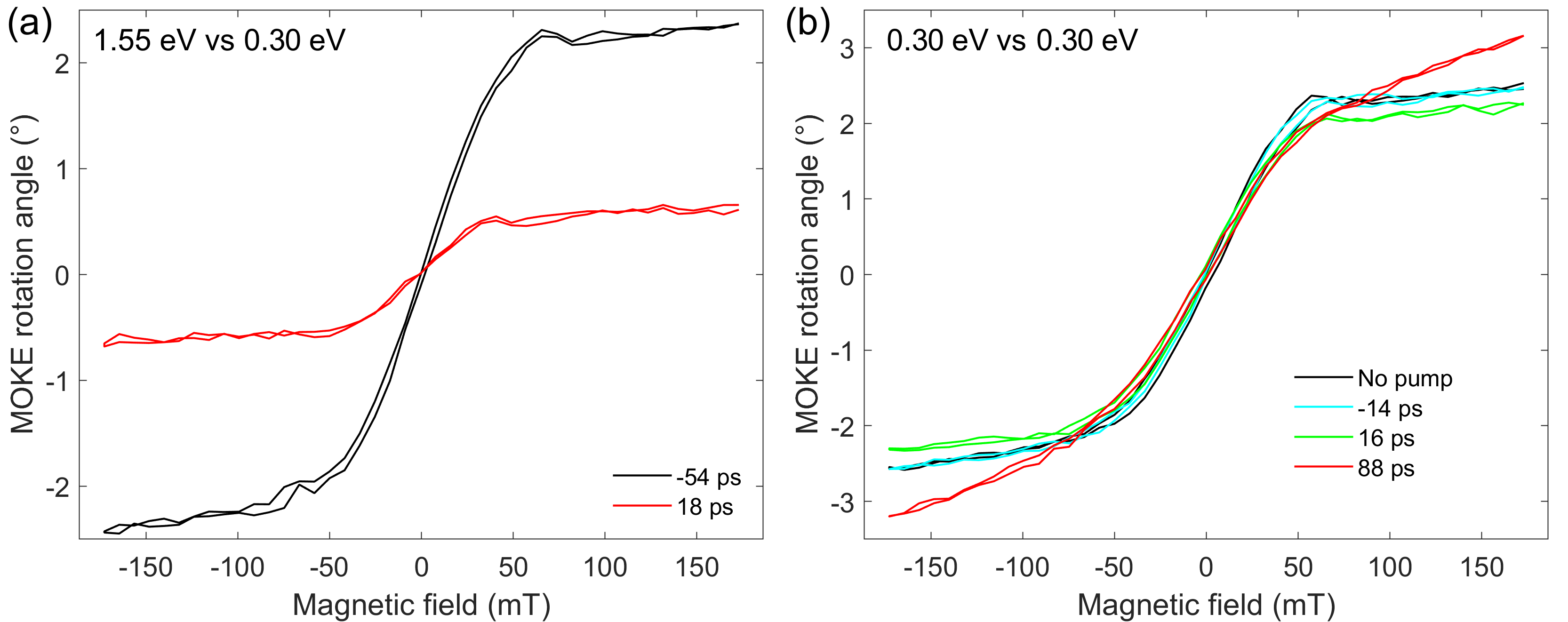}

\caption{Pumped MOKE rotation curved at different time delays for 0.30 eV probe photon energy, using (a) 1.55 eV (5.2 mJ/cm\textsuperscript{2}) and (b) 0.30 eV (10.0 mJ/cm\textsuperscript{2}) pump photon energies. All the measurements were acquired at $T$=75 K.}
\label{sfig7}
\end{figure}

To complement the TR-MOKE data presented in the Main Text, we here report `pumped', \textit{i.e.}, acquired in presence of the pump pulses, rotation hysteresis curves taken at different time delays for the various wavelength combinations employed. In Figs.
\ref{sfig6} and \ref{sfig7} we show the data obtained at the highest employed fluences in the Main Text. The corresponding time traces can be found in Figs. 2 and 3.

We observe that using a 3.10 eV probe (Fig.
\ref{sfig6}), every pump photon energy leads always to a reduction of the amplitude of the separation of the two hysteresis levels after a few tens of picoseconds.

Regarding the data based on the 0.30 eV probe, a similar behavior occurs for the 1.55 eV pump (Fig.
\ref{sfig7}(a) at 5.9 mJ/cm\textsuperscript{2}). However, for intermediate fluences (Fig. 3(e), 1.5-3.0 mJ/cm\textsuperscript{2}), we notice that the rotation dynamics changes sign after a few tens of picoseconds and persists at longer time delays (Fig. 3(f)). In the degenerate 0.30 eV setting,  this occurs still at the highest employed fluence (10.7 mJ/cm\textsuperscript{2}), whose corresponding `pumped' hysteresis is reported in Fig. \ref{sfig7}(b). Here we observe an increase in the rotation angle at the largest magnetic field moduli, with a `bending' of the hysteresis curve tails. In terms of excitation density, however, we remark that at 0.30 eV, the penetration depth is $\approx$2
times longer than the 1.55 eV case.


\clearpage
\section{MOKE Ellipticity}
\begin{figure}[h]
\includegraphics[width=9 cm]{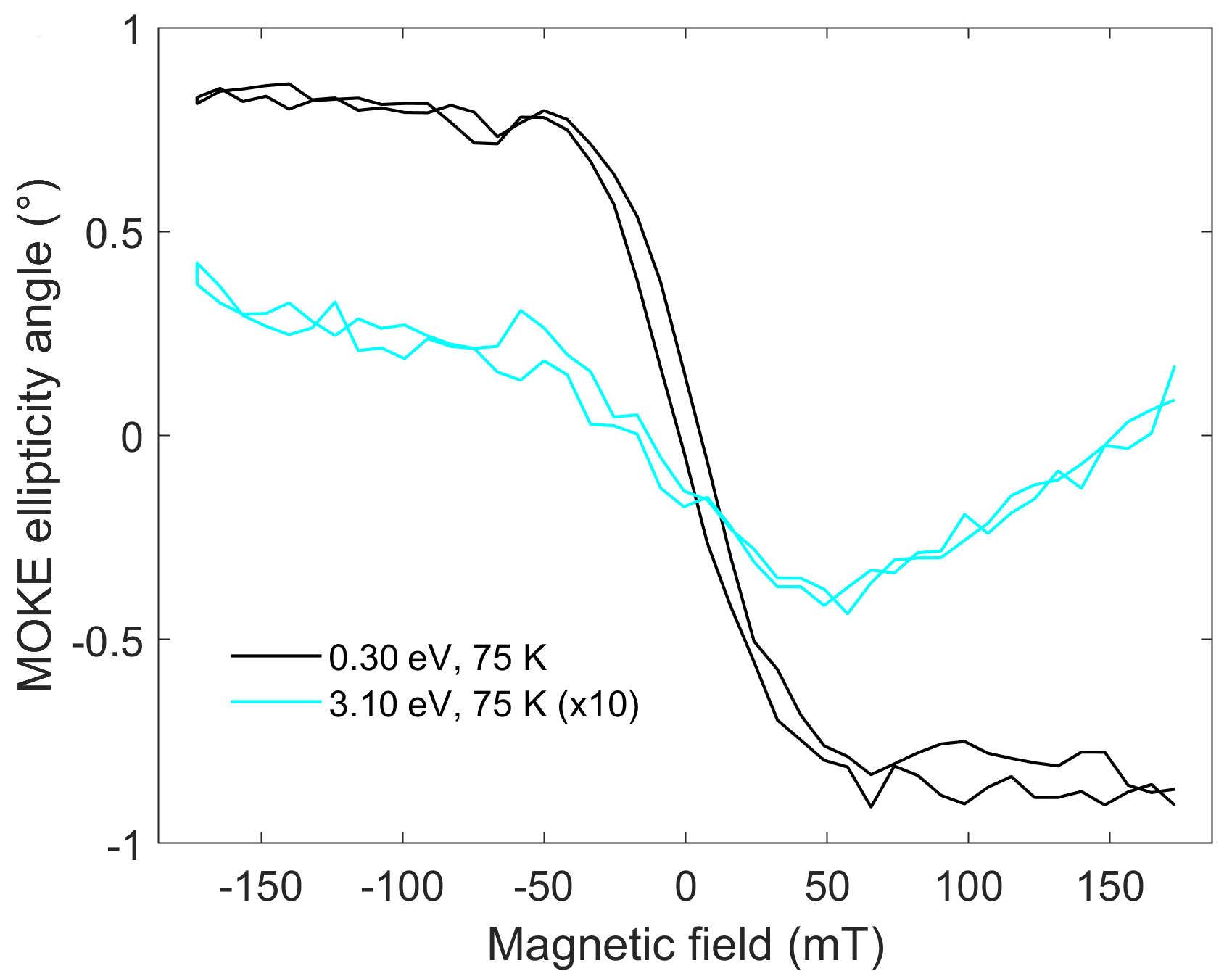}

\caption{Examples of ellipticity hysteresis curves for FeCr\textsubscript{2}S\textsubscript{4} in the ferrimagnetic phase.}

\label{sfig8}
\end{figure}
In Fig. \ref{sfig8}, we report examples of MOKE ellipticity curves collected at equilibrium on FeCr\textsubscript{2}S\textsubscript{4}.
For both 0.30 and 3.10 eV probe photon energies, the ellipticity values approach a zero-crossing, due to the energy dependence of the magneto-optical coefficients \cite{Ohgushi2005}. In this regard, non-uniformities in the sample can affect the shape of the energy dispersion of the coefficients leading to changes in the equilibrium ellipticity values by moving along the sample \cite{Klug2021}. Although we tried to avoid any movement as much as possible, refilling the cooling N\textsubscript{2} liquid and the necessity to ensure the spatial overlap between the pump and probe beams in the different wavelength configurations are operations which slightly modify the arrival of the beam at the sample. Therefore, we always refer the out-of-equilibrium changes to the equilibrium hysteresis acquired for each configuration.

\begin{figure}[h]
\includegraphics[width=\textwidth]{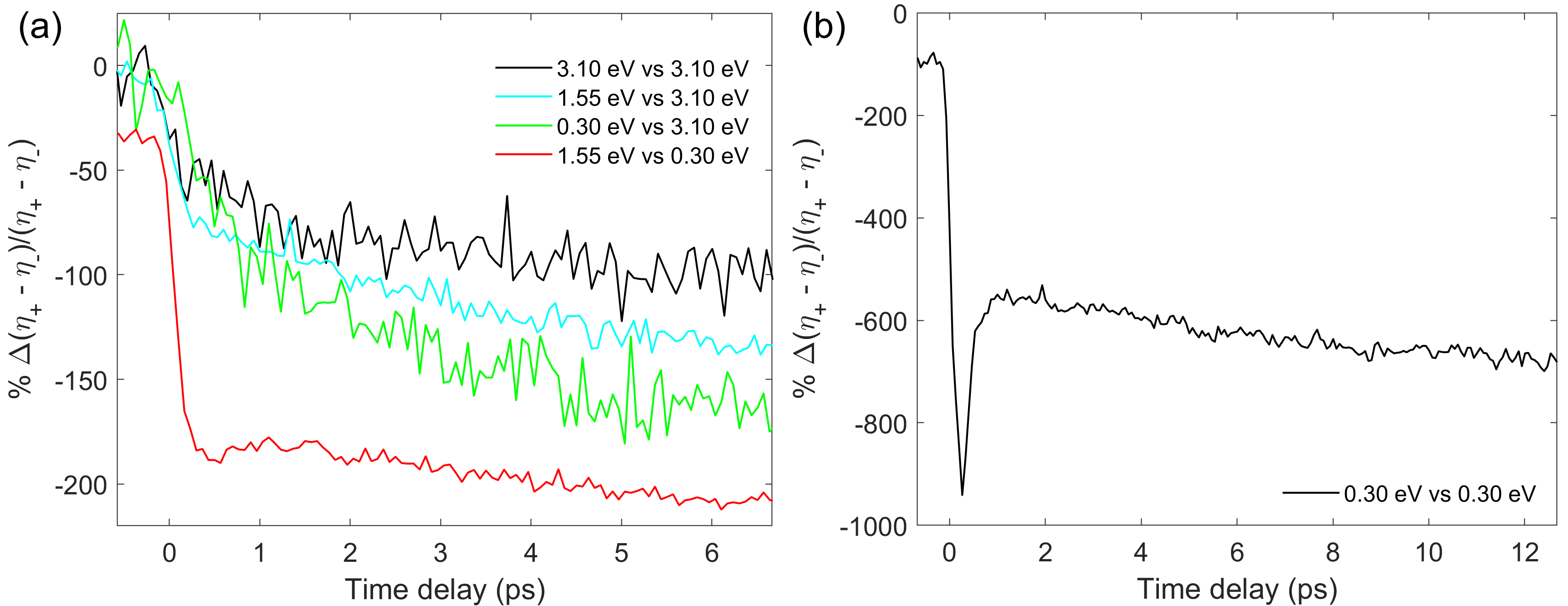}

\caption{(a),(b) TR-MOKE ellipticity dynamics for FeCr\textsubscript{2}S\textsubscript{4} in a short delay range for different wavelength configurations (pump vs probe photon energies). The pump fluences were the following: 3.10 eV vs 3.10 eV (1.9 mJ/cm\textsuperscript{2}), 1.55 eV vs 3.10 eV (5.0 mJ/cm\textsuperscript{2}), 0.30 eV vs 3.10 eV (6.4 mJ/cm\textsuperscript{2}), 1.55 eV vs 0.30 eV (5.2 mJ/cm\textsuperscript{2}) and 0.30 eV vs 0.30 eV (10.0 mJ/cm\textsuperscript{2}).  All the measurements were acquired at $T$=75 K.}
\label{sfig9}
\end{figure}

\begin{figure}[h]
\includegraphics[width=\textwidth]{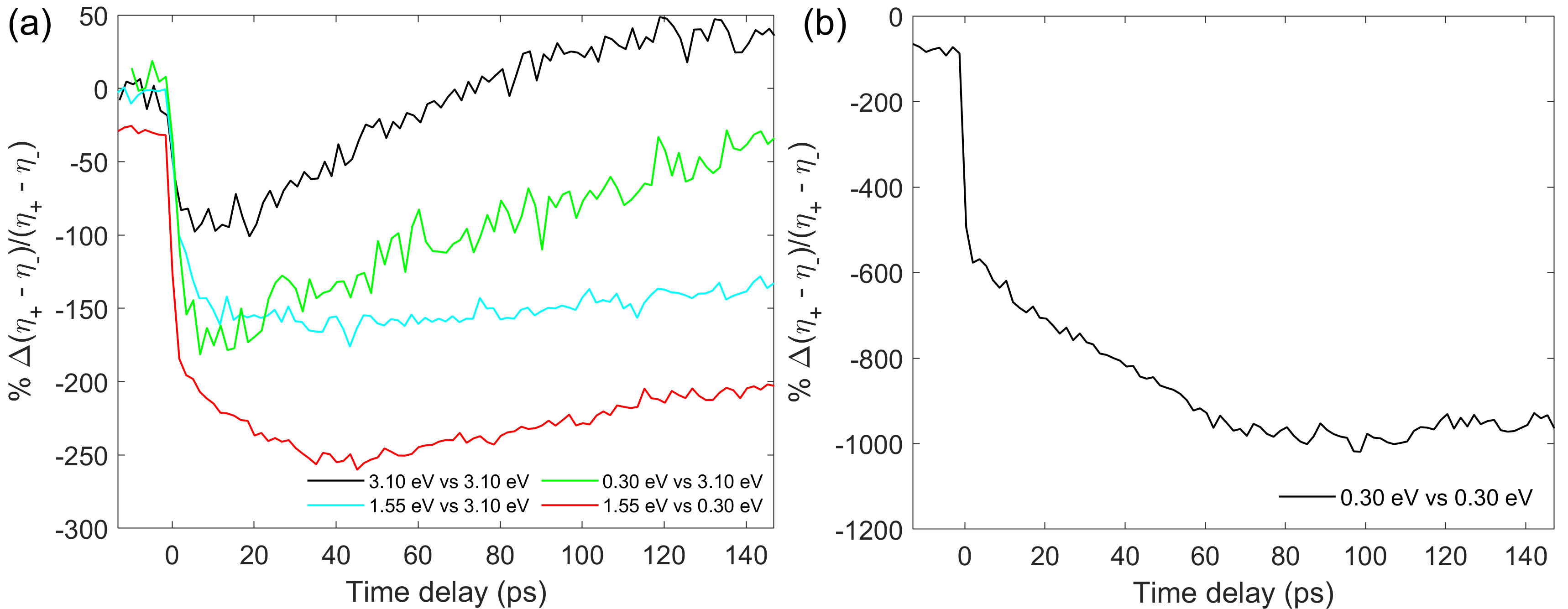}

\caption{(a),(b) TR-MOKE ellipticity dynamics for FeCr\textsubscript{2}S\textsubscript{4} in an intermediate delay range for different wavelength configurations (pump vs probe photon energies). The pump fluences were the following: 3.10 eV vs 3.10 eV (1.9 mJ/cm\textsuperscript{2}), 1.55 eV vs 3.10 eV (5.0 mJ/cm\textsuperscript{2}), 0.30 eV vs 3.10 eV (6.4 mJ/cm\textsuperscript{2}), 1.55 eV vs 0.30 eV (5.2 mJ/cm\textsuperscript{2}) and 0.30 eV vs 0.30 eV (10.0 mJ/cm\textsuperscript{2}). All the measurements were acquired at $T$=75 K.}
\label{sfig10}
\end{figure}

\begin{figure}[h]
\includegraphics[width=\textwidth]{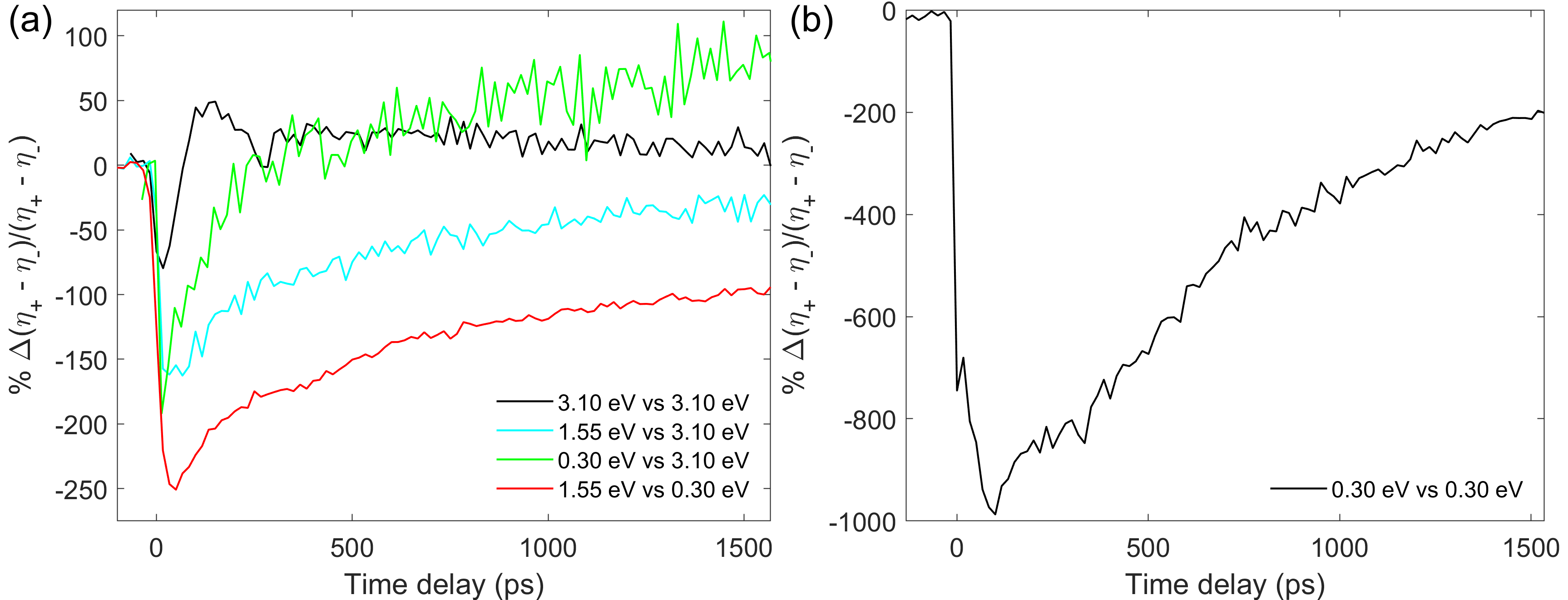}

\caption{(a),(b) TR-MOKE ellipticity dynamics for FeCr\textsubscript{2}S\textsubscript{4} in a long delay range for different wavelength configurations (pump vs probe photon energies). The pump fluences were the following: 3.10 eV vs 3.10 eV (1.9 mJ/cm\textsuperscript{2}), 1.55 eV vs 3.10 eV (5.0 mJ/cm\textsuperscript{2}), 0.30 eV vs 3.10 eV (6.4 mJ/cm\textsuperscript{2}), 1.55 eV vs 0.30 eV (5.2 mJ/cm\textsuperscript{2}) and 0.30 eV vs 0.30 eV (10.0 mJ/cm\textsuperscript{2}).  All the measurements were acquired at $T$=75 K.}
\label{sfig11}
\end{figure}

\begin{figure}[h]
\includegraphics[width=\textwidth]{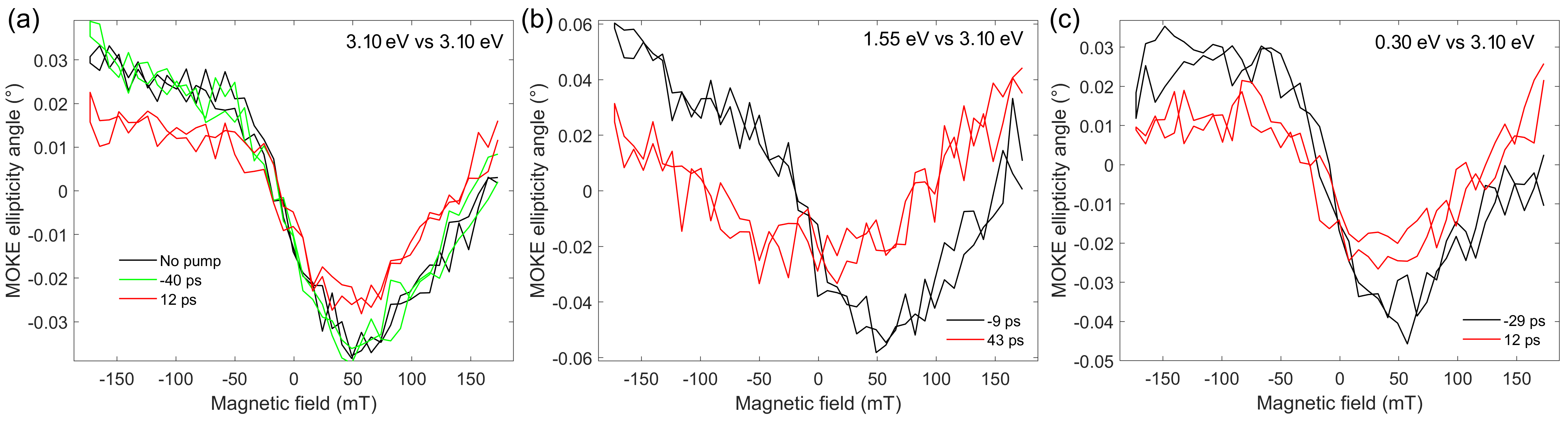}

\caption{Pumped MOKE ellipticity curved at different time delays for 3.10 eV probe photon energy, using (a) 3.10 eV (2.0 mJ/cm\textsuperscript{2}), (b) 1.55 eV (5.7 mJ/cm\textsuperscript{2}) and (c) 0.30 eV (6.4 mJ/cm\textsuperscript{2}) pump photon energies. All the measurements were acquired at $T$=75 K.}
\label{sfig12}
\end{figure}

\begin{figure}[h]
\includegraphics[width=\textwidth]{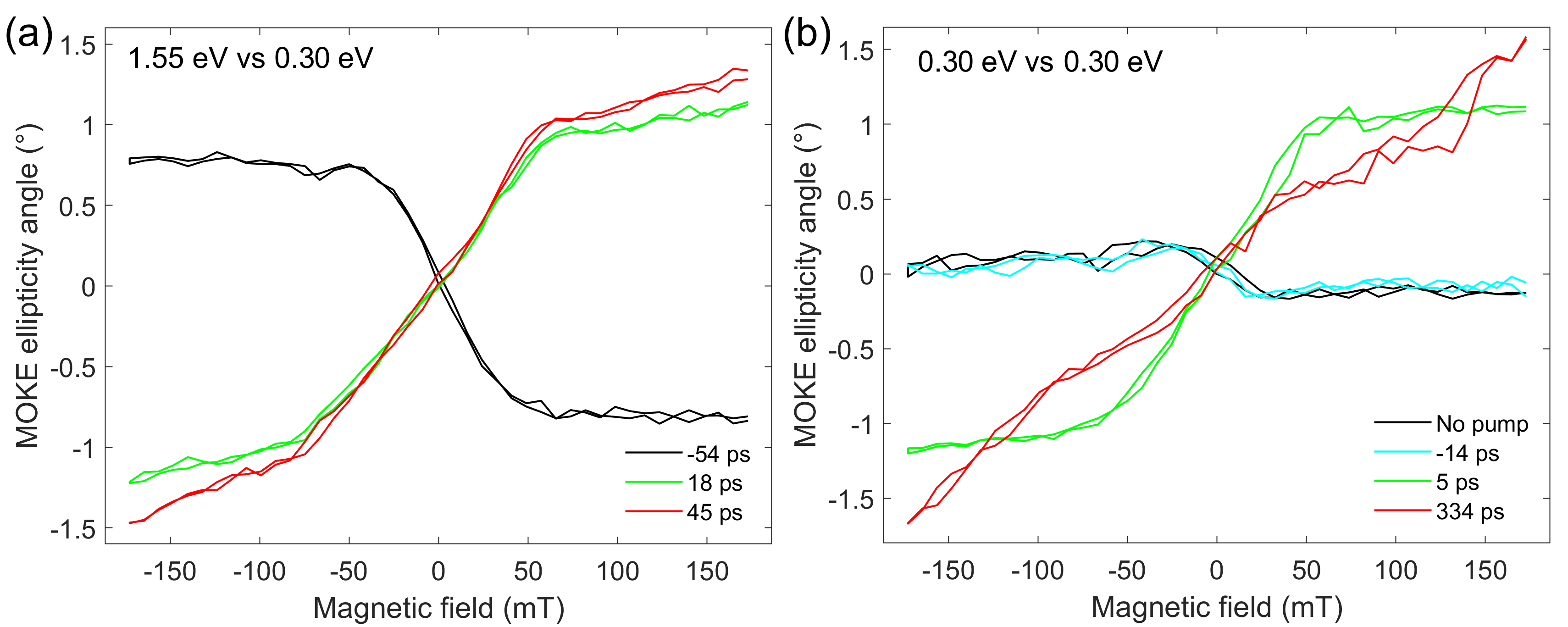}

\caption{Pumped MOKE ellipticity curved at different time delays for 0.30 eV probe photon energy, using (a) 1.55 eV (5.2 mJ/cm\textsuperscript{2}) and (b) 0.30 eV (10.0 mJ/cm\textsuperscript{2}) pump photon energies.  All the measurements were acquired at $T$=75 K.}
\label{sfig13}
\end{figure}

In Figs. \ref{sfig9}, \ref{sfig10} and \ref{sfig11}, we present the time-resolved ellipticity MOKE dynamics in three different time ranges at the highest fluences employed for each photon energy configuration for the rotation dynamics. As shown in Fig. \ref{sfig9}(a), we observe once again a slower dynamics for the resonant 0.30 eV pumping compared to the non-resonant 1.55 eV and 3.10 eV, excluding a non-magnetic response causing differences in the dynamics between rotation and ellipticity signals \cite{Koopmans2003, Checkhov2021}. The spike in the response near time-zero in Fig. \ref{sfig9}(b) is likely connected to the previously discussed coherent artifacts in the degenerate configuration.
We also collected at different time delays `pumped' hysteresis curves (Figs. \ref{sfig12} and \ref{sfig13}) showing the curves before and after the pump arrival, comparing each of the employed wavelength configurations.

Given that we are close to the zero-crossings of the magneto-optical coefficients, the relative out-of-equilibrium changes can be quite large, as it can be seen from the time traces.
Nevertheless, the observed dynamics share some features with the corresponding rotation traces in Figs. 2 and 3.

The results using a 3.10 eV probe, together with the `pumped' hysteresis curves in Fig. \ref{sfig12}, show a general reduction of the hysteresis curve amplitude, which is consistent with the rotation dynamics in Figs. 2, 3. We, however, notice values that exceed 100\% in relative variation for most of the recorded photon energies combinations. A similar situation was recorded for the 0.30 eV probe, where we observe a change in sign of the hysteresis curves for positive time delays (Fig. \ref{sfig13}). We believe this is caused a change of the magneto-optical coefficients due to the pump energy deposition. In fact, as shown in \cite{Ohgushi2005}, the rotation and ellipticity angles show a temperature dependence, which is not simply a global rescaling of the amplitudes. Together with it, a shift of the zero-crossing position occurs, which is indeed much more evident around 0.30 eV than at 3.10 eV. Therefore, we attribute this effect to a change in the magneto-optical coefficients, driven by a temperature increase in the sample.


\clearpage
\section{Time-resolved reflectivity}

The time-resolved reflectivity (\textDelta R/R, \textit{i.e.}, the ratio between the reflectivity change and the equilibrium reflectivity) was recorded removing the half- (or quarter-) wave plate, and Wollaston polarizer from the detection scheme, thus collecting all the reflected probe beam in a single diode.

\begin{figure}[h]
\includegraphics[width=\textwidth]{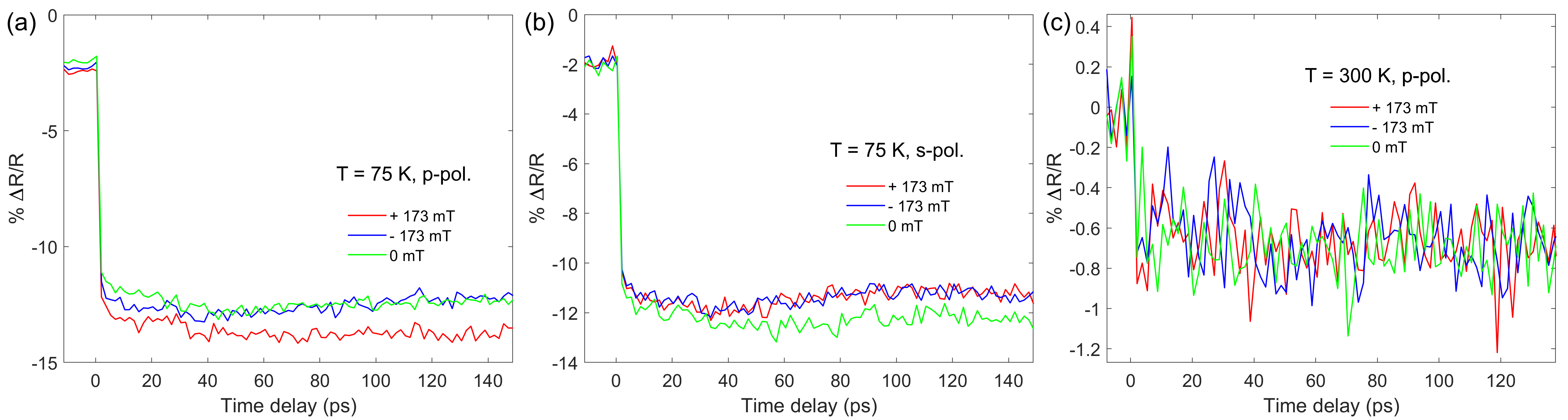}

\caption{Time-resolved reflectivity dynamics for FeCr\textsubscript{2}S\textsubscript{4} using a 1.55 eV pump and 0.30 eV probe under a 5.2 mJ/cm\textsuperscript{2} incident fluence. (a) p-polarized probe at $T$=75 K, (b) s-polarized probe at $T$=75 K and (c) p-polarized probe at $T$=300 K.}
\label{sfig14}
\end{figure}

Below the critical temperature, in the 1.55 eV pump and 0.30 eV probe configuration, we observe in Fig. \ref{sfig14} a clear influence of the magnetic field. In panel (a) (p-polarized probe), the positive magnetic field leads to a larger \textDelta R/R, while the negative and zero field traces are for the most part superimposed in the recorded time range. Differently, in panel (b) (s-polarized probe) the two curves of the sample under a magnetic field are superimposed with a smaller \textDelta R/R compared to the zero field case. 
We attribute this response, showing a difference between the direction of magnetic field with the same modulus, to a transversal component in our polar MOKE measurements. This can be present due to the small, albeit finite, angle of the two beams with respect to the surface normal (pump $2^{\circ}$ and probe $7^{\circ}$).
The initial offset is connected to a low-energy pre-pump pulse arriving 19 ps before $t$=0 coming from the pump beam, which, however, has a limited impact on the equilibrium state (Fig. 3(a),(c)). It was estimated to be about eight times smaller than the main pulse.

Above the critical temperature (panel (c)), under the same incident fluence, the \textDelta R/R becomes independent of the applied magnetic field. Moreover, the magnitude of the maximum reflectivity change decreases by one order of magnitude.

\bibliography{biblio}